\documentclass[twocolumns]{sdl2}

\usepackage{multirow}
\usepackage{multicol}
\usepackage{array}
\usepackage{tabularx}

\usepackage{stackengine}

\def\zR{z_{_R}}
\def\w{\mathrm{w}_{\0}}

\def\0{\mbox{\tiny $0$}}
\def\1{\mbox{\tiny $1$}}
\def\2{\mbox{\tiny $2$}}
\def\3{\mbox{\tiny $3$}}
\def\4{\mbox{\tiny $4$}}
\def\5{\mbox{\tiny $5$}}
\def\6{\mbox{\tiny $6$}}
\def\7{\mbox{\tiny $7$}}
\def\8{\mbox{\tiny $8$}}
\def\9{\mbox{\tiny $9$}}

\def\intxy{\int\mathrm{d} \kx   
\int \mathrm{d} \ky\,\,}

\def\intxyr{\int\mathrm{d} \xre   
\int \mathrm{d} y\,\,}

\def\in{^{^{\mathrm{[inc]}}}}
\def\re{^{^{\mathrm{[ref]}}}}
\def\tr{^{^{\mathrm{[tra]}}}}
\def\pol{_{_{\mathrm{pol}}}}
\def\tm{_{_{\mathrm{tm}}}}
\def\te{_{_{\mathrm{te}}}}

\def\wo{\mathrm{w}_{\0}}
\def\wos{\mathrm{w}_{\0}^{\2}}
\def\woc{\mathrm{w}_{\0}^{\3}}

\def\wz{\mathrm{w}(z)}
\def\wzs{\mathrm{w}^{\2}(z)}

\def\w{\mathrm{w}}

\def\s{^{\2}}
\def\c{^{\3}}
\def\q{^{^2}}

\def\G{G(\kx,\ky)}

\def\Gref{G\pol\re(\kx,\ky)}
\def\Gtra{G\pol\tr(\kx,\ky)}

\def\Einc{E\in(x,y,z)}
\def\Eo{E_{\0}}
\def\Iinc{I\in(x,y,z)}
\def\Io{I_{\0}}

\def\Eref{E\re\pol(\xre,y,\zre)}
\def\Iref{I\re\pol(\xre,y,\zre)}

\def\Irefo{I\re\tm(\xre,0,\zre)}

\def\Etra{E\tr\pol(\xtr,y,\ztr)}
\def\Itra{I\tr\pol(\xtr,y,\ztr)}

\def\xs{x_{*}}

\def\zs{z_{*}}

\def\xt{\widetilde{x}}

\def\zt{\widetilde{z}}

\def\xre{x_{_{\,\mathrm{ref}}}}
\def\zre{z_{_{\,\mathrm{ref}}}}

\def\xtr{x_{_{\,\mathrm{tra}}}}
\def\ztr{z_{_{\,\mathrm{tra}}}}

\def\xttr{\widetilde{x}_{_{\,\mathrm{tra}}}}

\def\kzs{k_{_{z_{*}}}}
\def\qzs{q_{_{z_{*}}}}
\def\qxs{q_{_{x_{*}}}}

\def\kx{k_{_x}}
\def\ky{k_{_y}}
\def\kz{k_{_z}}

\def\kxt{k_{_{\tilde{x}}}}

\def\kzt{k_{_{\tilde{z}}}}
\def\qzt{q_{_{\tilde{z}}}}
\def\qxt{q_{_{\tilde{x}}}}

\def\tbre{\theta_{_{\mathrm{exB}}}}
\def\tbri{\theta_{_{\mathrm{inB}}}}

\def\bex{_{_{\mathrm{exB}}}}
\def\bin{_{_{\mathrm{inB}}}}

\def\cri{_{_{\mathrm{cri}}}}

\def\sn{_{_{\mathrm{Snell}}}}

\def\ua{^{^{[1]}}}
\def\ub{^{^{[2]}}}
\def\uc{^{^{[3]}}}

\def\figureone{
\WideFigureSideCaption{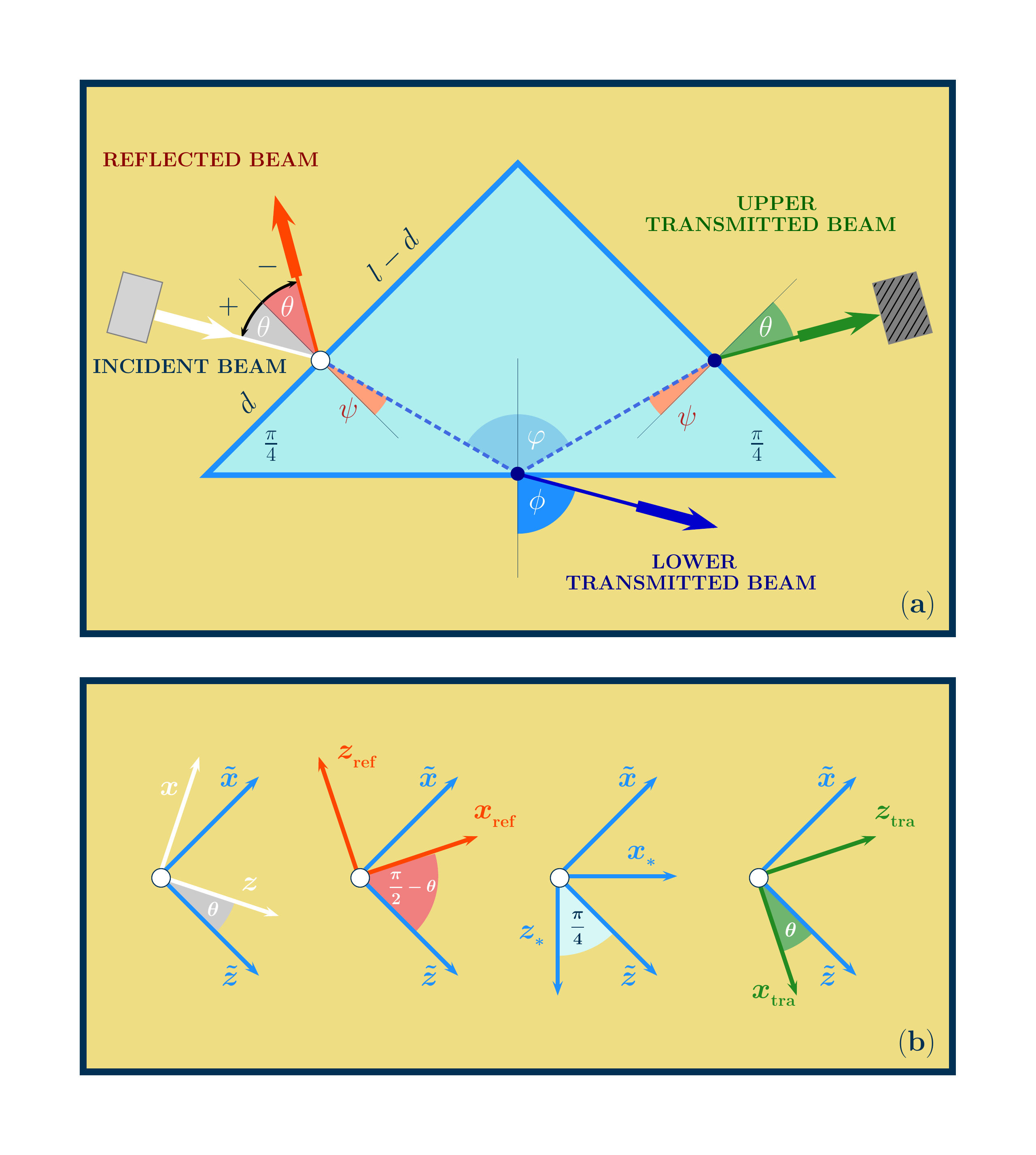}{The optical system used in this paper and the coordinate systems of the incident, the lower and upper transmitted beams. The $\widetilde{z}$ and $\widetilde{x}$ axes are respectively  the normals to the incident air/dielectric  interface and the upper dielectric/air interface. The $z_*$ axis  is the normal to the lower dielectric/air interface. The angles $\psi$,  $\varphi$, and $\phi$  contain an implicit dependence on the incidence angle $\theta$ and the refractive index $n$: $\sin\theta=n\,\sin\psi$,   $\varphi=\psi+\pi/4$,   and 
$n\,\sin\varphi=\sin\phi$.        
\label{fig1}}
}

\def\figuretwo{
\WideFigureSideCaption{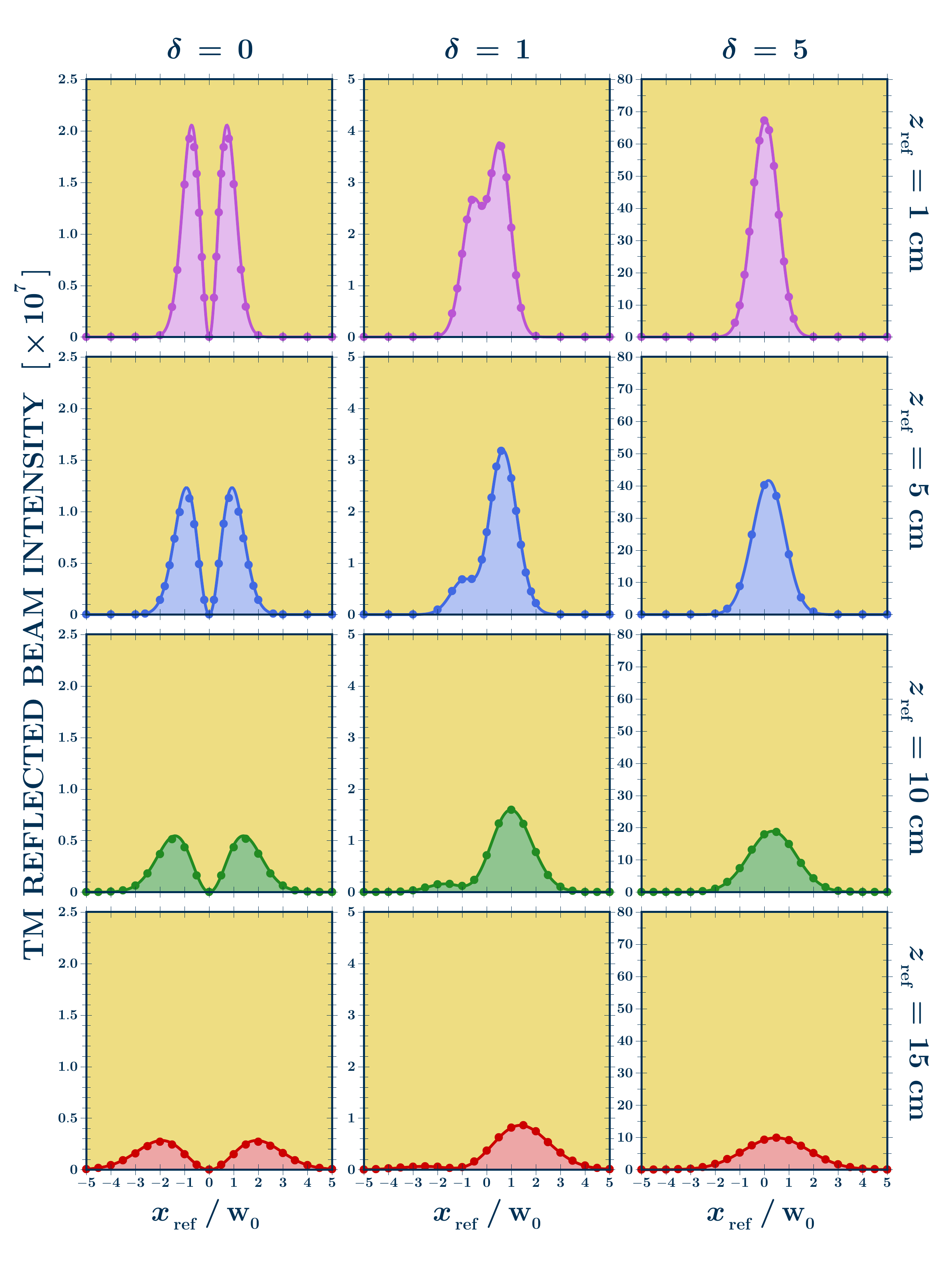}{Field intensity profiles of the reflected beam, $\Irefo/\Io$, obtained for an incident  Gaussian laser with  $\wo=100\,\mu\mathrm{m}$ and $\lambda=532\,\mathrm{nm} $. The plots refer to different incidence angles, 
$\tbre + \delta\,/\,k\wo$,  and axial distances and a BK7 prism ($n=1.5195$). The analytic curves (continuous lines) and numeric data (plots) show an excellent agreement. At the Brewster angle, we find the phenomenon of the double peak intensity.  For the laser parameters used,  the incidence $\delta=1$ corresponds  to  $\tbre + 0.05^{o}$.   \label{fig2}}}

\def\figurethree{
\WideFigureSideCaption{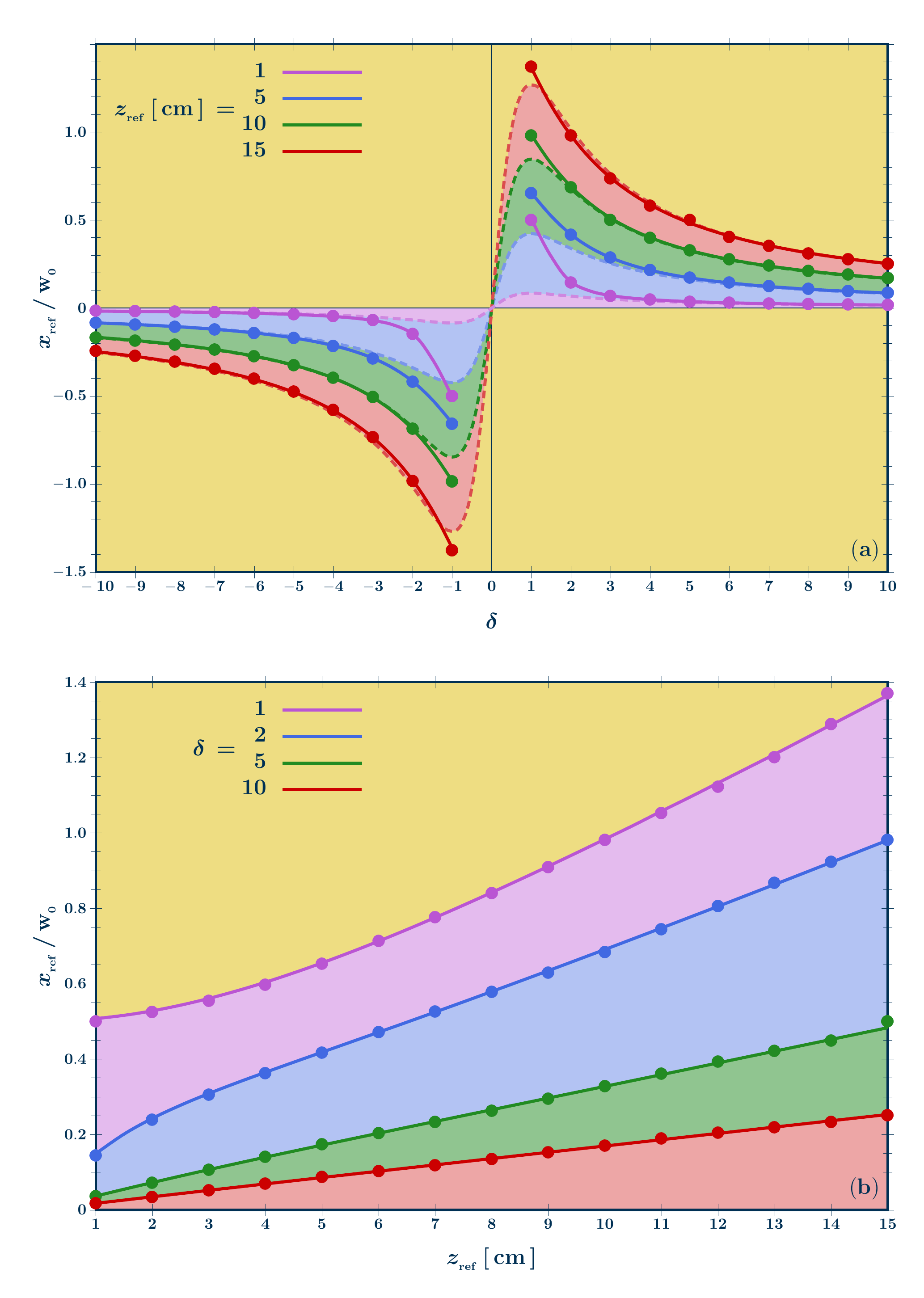}{
The solutions of the cubic equation for transverse magnetic waves and incidence near the Brewster angle (continuous lines)  are plotted as a function of $\delta$ for different values of $\zre$ (upper plots) and as a function  of  $\zre$ for different values of $\delta$ (lower plots). The mean value calculation and numerical data respectively 
appear as  dashed lines and dots. The beam parameters are $\w=100\,\mu\mathrm{m}$ and $\lambda=0.532\,\mathrm{nm}$
and the refractive index $n=1.5195$. The breaking of symmetry is evident when we approach the external Brewster angle and decrease the axial distance.
\label{fig3}}}

\def\figurefour{
\WideFigureSideCaption{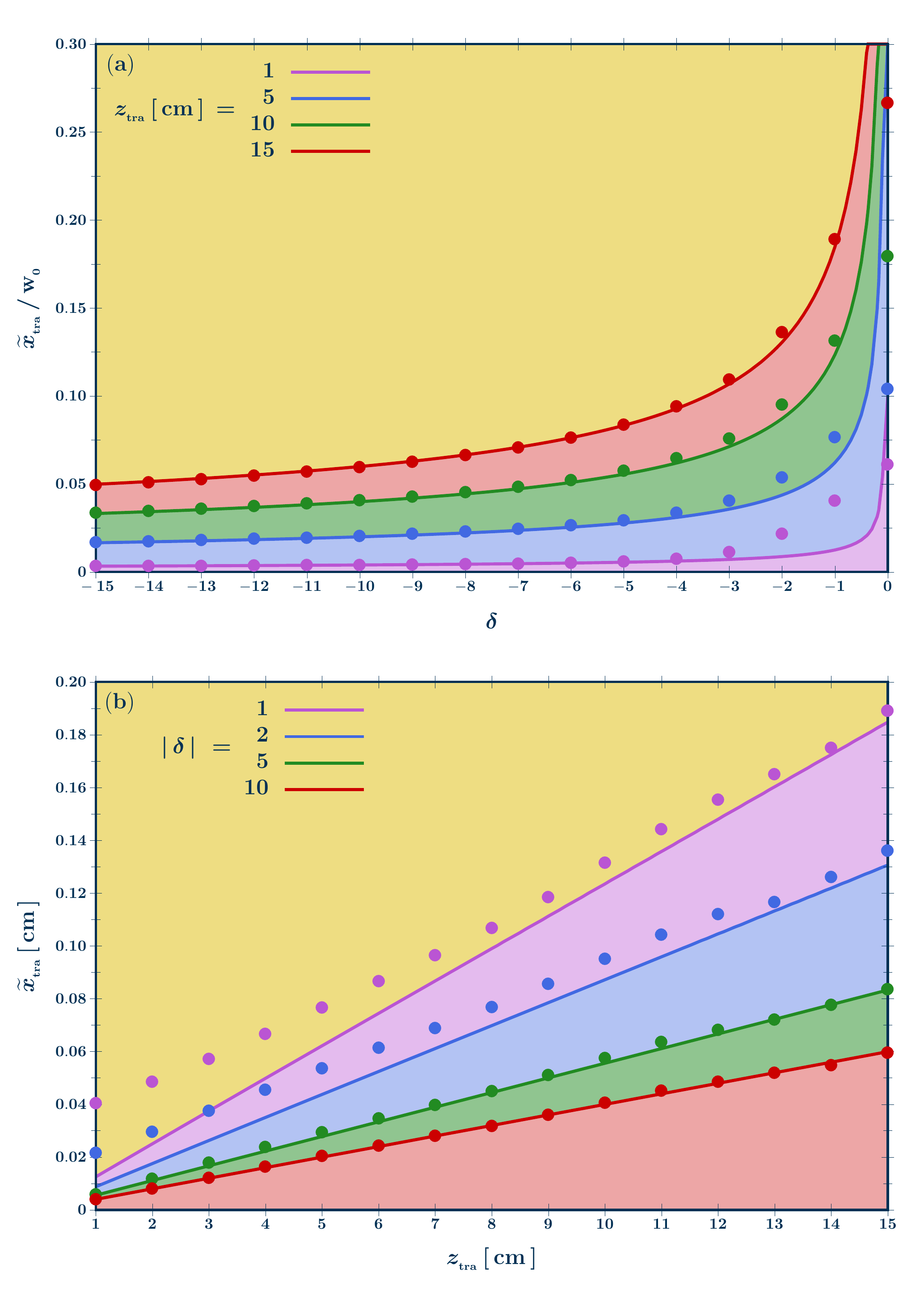}{Angular deviations for transverse magnetic beam (upper) transmitted through  a dielectric prism for incidence approaching the critical angle. In (a) the shift of peak as a function of the incidence angle for different axial distances. In (b) the shift is plotted as a function of the axial distance
for different values of the incidence angle. Analytic and numerical data show agreement for $\delta\leq-\,5$.
\label{fig4}}}

\def\figurefive{
\WideFigureSideCaption{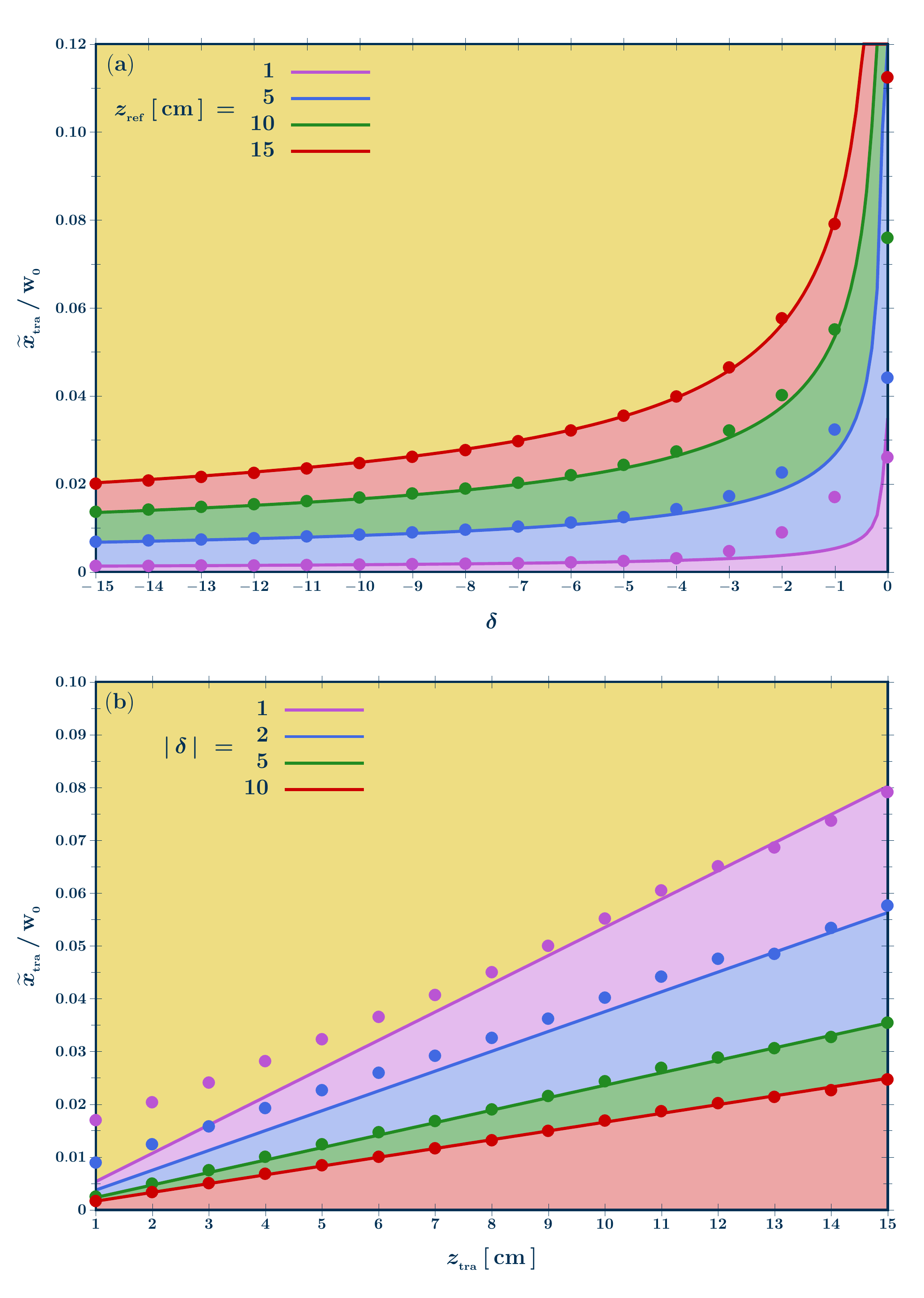}{The same of Figure 4 for transverse electric beams.
\label{fig5}}}

\journal{\shadowtext{\textbf{\color{DarkRed} European Physical Journal PLus}} \, \textbf{136}, 507-20 (2021).}

\titlelines{4}
\title{Angular deviations: From a cubic equation to\\ a  universal closed formula  to determine\\  the  peak position  of  reflected and\\ (upper) transmitted beams}

\imgbgabstract{Angular deviations and lateral displacements are optical effects widely investigated in literature. In this paper, by using the Taylor expansion of the Fresnel coefficients,  we obtain an analytic expression  for the beam reflected by and (upper) transmitted through a dielectric prism. These  analytical approximations lead to a cubic equation which allows to  determine the angular deviations of the optical beams. Near the Brewster angles, under specific conditions,  we obtain a universal formulation  for the cubic equation. Its explicit solution  determines
the peak position of the reflected and (upper) transmitted beams. The universal solution could be of great utility in future  experimental implementations. The analytic results show an excellent agreement  with the numerical calculation  and the analytic expressions given for the reflected and (upper) transmitted beams  should play an important role in the weak measurements analysis.}

\author{
\names{Stefano De Leo \textsuperscript{1,3} and Alessia Stefano\textsuperscript{2}}
\affiliation{\textsuperscript{1}Department of Applied Mathematics, State University of Campinas, Brazil}{\textsuperscript{2} Department of Mathematics and Physics, University of Salento, Italy}
\email{\textsuperscript{3} deleo@unicamp.br}
}

\begin{document}

\sdlmaketitle

\section{Introduction}
%%%%%%%%%%%%%%%%%%%%%%
%%%%%%%%%%%%%%%%%%%%%%
\figureone
%%%%%%%%%%%%%%%%%%%%%%
%%%%%%%%%%%%%%%%%%%%%%%

%Correction of Reflection law \\
%Breaking of simmetry\\
%Double peak in Brewster Angle\\
%Amplifications in Brewster Angle and in Critical Angle\\
%Analitycal expression: cubic equation\\
%Study of TM and TE component wuth Weak Measure\\

%Key words:\\
%Guassian distribution of optical beam, Gaussian beam\\
%Breaking of simmetry\\
%Fresnel coefficents\\
%Snell's law, refection law, geometrical optics\\

The optical path of a Gaussian beam reflected by or transmitted trough  a dielectric prism can easily be calculated by using  the laws of geometric optics\cite{born,saleh}. As the reflection and Snell laws surely represent a useful tool to describe the propagation of optical beams, they  cannot be used to predict or explain  phenomena like 
lateral displacements\cite{LD01,LD02,LD03,LD04,LD05,LD06,LD07,LD08,LD09,LD10,LD11,LD12,LD13,LD14,LD15a,LD15b,
LD16,LD17,LD18,LD19} and/or   angular deviations\cite{AD01,AD02,AD03,AD04,AD05,AD06,AD07,AD08,AD09,AD10,AD11,AD12}. 

Lateral displacements, known as Goos-H\"anchen shifts in honour of the German researchers that in 1947  realized, for transverse electric waves, the first experiment of this phenomenon\cite{LD01}, can theoretically be explained by using the stationary phase method as done by Artmann in 1948 \cite{LD02}. The additional phase, which characterizes the Fresnel coefficients in the case of  total reflection, is indeed responsible for the lateral displacement.
The Artmann formula provides an accurate estimate of the lateral displacement in terms 
of the light wavelength, refractive index, and incidence angle. It is polarization dependent and, consequently, implies  different lateral shifts for transverse electric and magnetic waves.  After the Artmann work, the German researchers tested the theoretical Artmann prediction by repeating the experiment  for transverse magnetic waves \cite{LD05}.  Due to the fact that the shifts are proportional to the beam wavelength,  Goos and H\"anchen used an  an elongated dielectric block to amplify the effect. For an optical  beam transmitted through a dielectric prism, lateral displacements appear in the upper transmitted beam when the angle, $\theta$,    that the incident beam forms with the first air/dielectric interface is greater than the critical angle, $\varphi\cri=\arcsin(1/n)$, see Figure 1.  In this case, it is common to use weak measurements to amplify the lateral displacements \cite{LD15a,LD18}.   In the region of \textit{pure} total reflection, the prediction of the Artmann formula were widely  confirmed in the experiments \cite{LD15a,LD15b,LD18,LD19}. Nevertheless, for incidence at critical angle, the Artmann lateral displacement  tends to infinity and numerical calculations are needed to reproduce the experimental data. In 1971\cite{LD08}, Horowitz and Tamir,  by using a Fresnel approximation in the Gaussian integral form of the optical beam, derived the lateral displacement for angles of incidence that are arbitrarily close to the critical one. They proposed a formula in terms of parabolic-cylinder (Weber) functions which reproduced the classical one  for large beam widths and for incidence angles that are not too close to the critical angle. As the incidence angle approaches the critical angle, the beam shift approaches a constant value that is dependent on the beam width\cite{LD08,LD16}. They  also found that the maximum lateral displacement occurred at an angle that is slightly larger than the critical one.  In 2016 \cite{LD17}, a closed formula in terms of modified Bessel functions of the first kind was obtained by studying the angular distribution of the optical beam. The new formula   allowed to understand how the breaking of symmetry in the angular distribution causes a difference between experimental measurements done by calculating  the maximum and the mean value of the beam intensity. This study also showed the lateral displacement dependence on the angular distribution shape of the incoming beam. The theoretical prediction  were confirmed by the experiments described in\cite{LD18,LD19}.

As described before, for \textit{pure} total reflection, the additional phase which appears in the Fresnel coefficients is responsible for lateral displacements. In the case of \textit{pure} partial reflection, the Fresnel coefficients are real, so the optical beam does not present any lateral shifts. Nevertheless, the breaking of symmetry in the Fresnel coefficients generates angular deviations.   In 1985\cite{AD03}, Chan and Tamir analysed the angular deviation  in the region around the Brewster angle. They used  a mathematical method  similar to the one of ref.\cite{LD08} offering an interesting presentation of the phenomenon near the Brewster angle, where the reflected beam, as we shall see explicitly later,  is so deformed in comparison to the incident one that the concept of angular deviation lacks any meaning. Few years later\cite{AD04}, the authors also discussed the angular deviations near the critical angle reviewing previous works\cite{AD01,AD02}.  In 2009 \cite{AD08,AD09},  Aiello and Woerdman revisited the topic of angular deviations in  the Brewster region\cite{AD05}, calculating the mean value of the  angular deviations from geometric optics. A closed form  expression for angular deviations of the Snell law in the case of maximal symmetry breaking at critical angle were obtained in \cite{AD11}. As it is happens for lateral displacements the weak measurement technique can be used to amplify such deviations\cite{AD12}. The reader can find a tutorial review of lateral shifts and angular deviations of Gaussian beams in \cite{AD13}. It is important to observe that for an optical beam reflect by and transmitted through a dielectric prism, we have \textit{three} Brewster angles: Two for the reflected beam,  $\theta\bex=\pm\arctan[n]$, and one for the upper transmitted one, 
$\varphi\bin=\arctan[1/n]$, see Figure 1.

For incidence near the critical angle, under certain conditions, angular deviations and lateral displacement act simultaneously generating the effect of axial dependence of the Goos-H\"anchen shift  and the oscillatory behaviour of light\cite{osc}. This intriguing phenomenon was recently confirmed experimentally\cite{ol1,ol2}  

The subject matter of our investigation is the study of angular deviations in the Brewster
regions for the optical beams reflected by and transmitted through a dielectric prism and in the critical region 
for the (upper) transmitted beam.   The paper is structured as follows. Section II introduces  the formalism, the free Gaussian beam, the optical system, and the Fresnel coefficients. Section III contains the  analytical approximation  of the reflected beam  from which we obtain  the cubic equation for the peak.  In this Section, for incidence near  the external Brewster angles and for an axial distance greater than the Rayleigh radius,  we obtain, for the angular deviations,  a universal formula to determine the peak of the beam. This allows to compare
the mean value and maximum analysis of angular deviations. The analytical predictions show an excellent agreement   with numerical calculation. In Section IV, we discuss the (upper) transmitted beam and the amplification near the internal  Brewster angle and the critical one.  Conclusions  and further investigations appear in the last Section.

\section{Spatial optical phases}
%%%%%%%%%%%%%%%%%%%%%%
%%%%%%%%%%%%%%%%%%%%%%
\figuretwo
%%%%%%%%%%%%%%%%%%%%%%
%%%%%%%%%%%%%%%%%%%%%%%
The propagation of a Gaussian beam which moves from the source to the left air/dielectric interface
of a dielectric prism, see Figure 1, is described by the following integral
\begin{eqnarray}
\label{inc}
\Einc &=& \Eo \intxy \G \,\,\times \nonumber \\ 
& & \exp\left[\,i\,(\,\kx x\,+\,\ky y\, +\, \kz z\,)\,\right]\,\,,
\end{eqnarray}
where
\begin{equation*}
\G\, = \,\wos\, \exp\left[\,-\,\left(\,\kx\s + \ky\s\,\right)\,\wos\,/\,4\,\right]\,/\,4\,\pi\,\,,
\end{equation*}
is the Gaussian wave number distribution, $\wo$ is the waist radius, and 
\[
\kz = \sqrt{k\s - \kx\s - \ky\s}\;\approx \;  k - (\kx\s + \ky\s)\,/\,2\, k
\]
is a function of $\kx$  and $\ky$ and contains the wavelength $\lambda$ $(k=2\pi/\lambda)$. 
By using the previous approximation, valid  for  $\wo\geq\lambda$, the integral in  Eq.\,(\ref{inc}) can be analytically solved 
\begin{equation}
\label{Einc}
\Einc =  \frac{\Eo\,\, e^{i  k z}}{1 +  i\, z/\zR}\;\exp \left[ - \frac{x\s +\, y\s}{\wos 
\left( 1 +  i\, z / \zR \right) }\right]
\end{equation}
where $\zR= k\, \wos/2$ is  the Rayleigh length. The intensity of the incident beam is then given by 
\begin{equation}
\label{Iinc}
\Iinc = \Io\,\frac{\wos}{\wzs} \,\exp{\left[ -\, 2 \,\frac{x\s + y\s}{\wzs}\right]} 
\end{equation}
where $\Io=|\Eo|\q$ and $\wz = \wo \sqrt{1 + (z / \zR)\q}$ is the radius at which  the field 
intensity falls to $1/e\s$.

In order to obtain the integral form of the reflected and (upper) transmitted beam, we first calculate the respective spatial optical phases. This will be done by introducing  the appropriate coordinate systems, see Figure 1.   For the left (air/dielectric) interface and the right (dielectric/air) one we have the $(\xt,\zt)$ system,
for the lower (dielectric/air interface) the $(\xs,\zs)$ system, and, finally, for the reflected and (upper) transmitted beam we respectively have the $(\xre,\zre)$   and  $(\xtr,\ztr)$ coordinates.

By a $\theta$  rotation, the spatial phase of the incident beam can be rewritten in terms of the left (air/dielectric)  interface system as follows 
\[\kx\,x\,+\,\ky\,y\,+\,\kz\,z\,=\,\kxt\,\xt\,+\,\ky\,y\,+\,\kzt\,\zt\,\,.\]
The spatial phase of the reflected beam is then given by
\[\kxt\,\xt\,+\,\ky\,y\,-\,\kzt\,\zt\,\,.  \]
By using the matrix algebra, we can rewrite
\begin{eqnarray*}
\left(\,\kxt\,\,\,\,\, \kzt\,\right)\,
\left(\begin{array}{r} \xt\\ -\,\zt\end{array}\right)\,\,,
\end{eqnarray*}
in terms of $\kx$, $\ky$, and the proper coordinate system of the reflected beam   
\begin{eqnarray*}
\left(\,\kx\,\,\,\,\, \kz\,\right)\,
\left(\begin{array}{rr} \cos\theta & -\,\sin\theta\\ \sin\theta & \cos\theta\end{array}\right)\,
\left(\begin{array}{rr} 1 & 0\\ 0 & -\,1 \end{array}\right) &\times\\
\left(\begin{array}{rr} \sin\theta & \cos\theta\\ -\,\cos\theta & \sin\theta\end{array}\right)\left(\begin{array}{c} \zre\\ \xre\end{array}\right)& = \\
\left(\,\kx\,\,\,\,\, \kz\,\right)\,\left(\begin{array}{c} \xre\\ \zre\end{array}\right)&.
\end{eqnarray*}
The integral form of the reflected beam,
\begin{eqnarray}
\label{Eref}
\Eref &=& \Eo \intxy \Gref  \,\,\times \nonumber \\ 
& &\hspace*{-1cm} \exp\left[\,i\,(\,\kx\,\xre\,+\,\ky\,y\,+\,\kz\,\zre\,)\,\right]\,\,,
\end{eqnarray}
is polarization dependent. This is due to the fact that  the wave number distribution,   
\[\Gref =R\pol(\kx,\ky)\,\G\,\,, \]
contains the Fresnel coefficients
\[R\pol(\kx,\ky) = \frac{a\pol\kzt -\qzt/a\pol}{a\pol\kzt+\qzt/a\pol}\,\,, \]
where
\[ \left\{\,a\te\,,\,a\tm\,\right\}\,=\,\{\,1\,,\,n\,\}\,\,,\] 
 with the subscript which indicate transverse electric (te) and magnetic (tm) waves.

The spatial phase of the transmitted beam at the left (air/dielectric) interface is given by
\[\qxt\,\xt\,+\,\ky\,y\,+\,\qzt\,\zt\,\,,\]
where 
\[\qxt=\kxt\,\,\,\,\,\mbox{and}\,\,\,\,\,\qzt= \sqrt{n\s k\s-\qxt\s-\ky\s}\,\,\]
 In terms of the coordinates system of the lower (dielectric/air) interface, the previous phase can be rewritten as
\[\qxs\,\xs\,+\,\ky\,y\,+\,\qzs\,\zs\]
which, after reflection at the lower (dielectric/air) interface, becomes 
\[\qxs\,\xs\,+\,\ky\,y\,-\,\qzs\,\zs\,\,.\]
As done for the reflected beam, by using the matrix algebra, we find
\begin{eqnarray*}
\left(\,\qxs\,\,\,\,\, \qzs\,\right)\,
\left(\begin{array}{r} \xs\\ -\,\zs\end{array}\right) & =\\
\frac{1}{\sqrt{2}}\,\left(\,\qxt\,\,\,\,\, \qzt\,\right)\,
\left(\begin{array}{rr} 1 & -\,1\\ 1& 1\end{array}\right)\,
\left(\begin{array}{rr} 1 & 0\\ 0 & -\,1 \end{array}\right)& \times\\
\frac{1}{\sqrt{2}}\,
\left(\begin{array}{rr} 1 & 1\\ -\,1 & 1\end{array}\right)\left(\begin{array}{c} \xt\\ \zt\end{array}\right)&=\\
\left(\,\qxt\,\,\,\,\, \qzt\,\right)\,
\left(\begin{array}{c} \zt\\ \xt\end{array}\right) & .
\end{eqnarray*}
From this spatial phase, observing that $\qxt=\kxt$ and that the normal to the right (dielectric/air) interface 
is $\xt$, we immediately obtain  the one of the transmitted beam   
\[\kxt\,\zt\,+\,\ky\,y\,+\,\kzt\,\xt\,\,.    \]
This phase can be rewritten in terms of $\kx$, $\ky$, and the proper system of the (upper) transmitted beam 
\begin{eqnarray*}
\left(\,\kx\,\,\,\,\, \kz\,\right)\,
\left(\begin{array}{rr} \cos\theta & -\,\sin\theta\\ \sin\theta & \cos\theta\end{array}\right)\,
\left(\begin{array}{rr} 0 & 1\\ 1& 0\end{array}\right)&\times\\
\left(\begin{array}{rr} \cos\theta & -\,\sin\theta\\ \sin\theta & \cos\theta\end{array}\right)
\left(\begin{array}{c} \ztr\\ \xtr\end{array}\right)&=\\
\left(\,\kx\,\,\,\,\, \kz\,\right)\,
\left(\begin{array}{c} \xtr\\ \ztr\end{array}\right) & .
\end{eqnarray*}
Finally,
\begin{eqnarray}
\label{Etra}
\Etra &=& \Eo \intxy  \Gtra \,\,\times \nonumber \\ 
& & \hspace*{-1cm} \exp\left[\,i\,(\,\kx\,\xtr\,+\,\ky\,y\,+\,\kz\,\ztr\,)\,\right]
\end{eqnarray}
where
\[\Gtra =T\pol(\kx,\ky)\,G(\kx,\ky) \]
and
\begin{eqnarray*}
T\pol(\kx,\ky)\, =\,  \frac{4\,\kzt\qzt}{(a\pol\kzt+\qzt/a\pol)\q}\,
\frac{\qzs/a\pol-a\pol\kzt  }{\qzs/a\pol+a\pol\kzs }\,\,\times \nonumber\\
  \exp\{\,i\,[\, \qzs d\,\sqrt{2}\,+\,(\,\qzt\,-\,\kzs\,)\,(\,\ell\,-\,d\,)\,]\,\}\,\,.
 \end{eqnarray*}
The phase which appears in the Fresnel coefficients is due to the fact that the lower  and right 
(dielectric/air) interfaces  do not coincide with the origin of the coordinate systems which we have fixed 
to the left (air/dielectric) interface. Experimentally, this means that the minimal waist is found at such an interface. Such a  phase is responsible  for the  optical geometrical path predicted by the Snell's law\cite{ajp}.

Numerical integrations of  Eqs.\,(\ref{Eref}) and (\ref{Etra}) can be used to study the propagation of the 
reflected and (upper) transmitted beam.  In particular, to analyse  angular deviations and lateral displacements. Nevertheless, analytical approximations for the optical beams are important to understand  the physical  parameters which characterize such phenomena and to determine the best  experimental configuration to observe angular deviations and lateral displacements. In this spirit, in next sections, we look for analytic expressions of  the reflected and (upper) transmitted beams. Such approximations, once tested by the numerical calculations, are then used to obtain a cubic equation from which it is possible  to determine  the angular deviations of the optical beams.

\section{The reflected beam}

In order to obtain an analytic expression for the reflected beam, we consider the first order Taylor expansion of the reflection coefficient, i.e.
\begin{equation}
\label{r1o}
R\pol(\kx,\kx)\,=\, R\pol(0,0)\, \left[\,1\,+\,\alpha\pol \,\kx/k\,\right]\,\,,
\end{equation}
where
\[
R\pol(0,0) = \frac{a\pol\cos\theta -n\cos\psi/a\pol}{a\pol\cos\theta+n\cos\psi/a\pol}\]
and 
\[
\begin{array}{lcl}
\alpha\te&=& 2\,\tan\psi\,\,,\\
 \alpha\tm &=&2\,\tan\psi\,/\, (\sin\s\psi\,-\,\cos\s{\theta})\,\,.
\end{array}
\]
Observing that the $\kx$ dependence  in (\ref{r1o})  can be substituted by  $-\,i\,\partial_{_{x}}$ in the integrand of the reflected beam,  Eq.\,(\ref{Eref}) becomes analytically solvable. The electric field of the reflected beam is
then approximated by
\begin{eqnarray}
\label{Eref2}
\Eref
&= &   \left[1 +\,i\,\, \frac{ \alpha\pol \xre +\zre}{\zR} \right]\times \nonumber \\
& & \hspace*{-1cm}\frac{R\pol ( 0 , 0 ) }{1 + i\, \zre / \zR}\,\, E\in ( \xre , y , \zre )\,\,.
\end{eqnarray}
Consequently, we find the following analytical expression for the field intensity 
\begin{eqnarray}
\label{Iref}
\Iref&=&  \left[1+ \left( \dfrac{\alpha\pol\xre +\zre}{\zR} \right)\s\, \right] \times \nonumber \\
& & \hspace*{-1.5cm} \dfrac{\wos}{\w\s(\zre)} \,R\pol^{^2} ( 0 , 0 )\,\,  I\in (\xre,y,\zre)\,\,.
\end{eqnarray}
This analytical approximation is checked by numerical calculations. The plots in Figure 2, done for different incidence angles, $\tbre + \delta\,/\,k\wo$,  and axial distances, show an excellent agreement between analytic curves and numeric data.

\subsection{The external Brewster angle}
As observed in the Introduction, we have to differentiate between the Brewster angles of the transverse magnetic  waves at the left (air/dielectric) and lower (dielectric/air) interfaces. So, we introduce the terminology of \textit{external} and \textit{internal} Brewster angles. The external Brewster angle refers to the reflection at the left (air/dielectric) interface and it is determined by $n\,\cos\theta=\cos\psi$. After simple algebraic manipulation, we find  
\begin{equation}
\label{extB}
\tan\tbre=\pm \,n\,\,.
\end{equation}
For incidence at Brewster  angle, the field  is given by  
\begin{eqnarray}
\label{IB}
I\bex\re(\xre,y,\zre) &=&   \left[f(n) \,\,\,\dfrac{\xre}{\zR}\, \dfrac{\wo}{\w(\zre)} \right]\q \times \nonumber \\   
& & I\in (\xre,y,\zre)\,\,,
\end{eqnarray}
where
\begin{eqnarray*}
f(n)  &=& \lim_{\,\,\,\,\,\theta\to\tbre} \alpha\tm R\tm ( 0 , 0 )\\ 
& = & \pm\,\,\frac{\sqrt{1+n\q}}{n\q}\, \lim_{\,\,\,\,\,\theta\to\tbre} \frac{n\,\cos\theta\,-\,\cos\psi}{\sin\q\psi\,-\,\cos\q\theta} =  \pm\,\,\frac{1-n^{^{4}}}{2\, n^{^{3}}}\,\,.
\end{eqnarray*}
The reflection coefficient $R\tm$ is zero, but it is compensated by the zero in the denominator of  $\alpha\tm$. 
From (\ref{IB}), we immediately calculate the position of the double peaks in the field intensity 
\[ x_{_\mathrm{ref}}^{^\mathrm{[max]}}\,=\,\pm\,\,\w(\zre)\,/\,\sqrt{2}\,\,. \] 
For a Gaussian beam with $\wo=100\,\mu\mathrm{m}$ and $\lambda=532\,\mathrm{nm}$, the  Rayleigh range is 5.9 cm.
Consequently, at the axial distances 1, 5, 10, and 15 cm, we find 
 
\[ x_{_\mathrm{ref}}^{^\mathrm{[max]}}\,/\,\wo=\,\pm\,\,\{\,0.72\,,\,0.93 \,,\,1.39\,,\, 1.93 \,\}   \,\,, \]
see Figure 2 for $\delta=0$. The maxima  of the field intensity are 
\[  \left[\dfrac{f(n)}{\sqrt{2}\,\pi} \,\,\dfrac{\lambda}{\wo}\right]^{^{2}} \times\,\,  \exp[-1]\,\,\times\,\, 
\left[\dfrac{\wo}{\w(\zre)} \right]^{^{2}}\,\,.  \]
For the Gaussian beam considered in this paper, we obtain the following  reduction factor  
\[  \left[\dfrac{f(1.5195)}{\sqrt{2}\,\pi} \,\,\dfrac{0.532}{100}\right]^{^{2}} \times \,\,\frac{1}{e} \,\,\approx \,\,2 \times\,10^{^{-7}}\] 
with respect to the intensity of the incident beam. The plots of Figure 2 show that by increasing $\delta$ we recover the symmetry of the incident beam and this allows to use the mean value calculation to estimate the angular deviations.

%%%%%%%%%%%%
%%%%%%%%%%%%
\figurethree
\figurefour
\figurefive
%%%%%%%%%%%%%%
%%%%%%%%%%%%%%

\subsection{The mean value calculation}
For a symmetric beam mean and maximum coincide so the angular deviations can be calculated by using
 \begin{equation}
\left<\xre\right>\pol =
\frac{\displaystyle{ \intxyr \xre \,\Iref}}{\displaystyle{ \intxyr\Iref}}\,\,.
\end{equation}
Observing that the intensity exponential  is an even  function in the transversal components, the $\xre$ linear 
term in Eq.\,(\ref{Iref}) is the only contribution to the numerator whereas the constant and quadratic terms are the ones which contribute to  the denominator,
\begin{eqnarray}
\left<\xre\right>\pol = \hspace{6cm}  &\\
\frac{\displaystyle{2\,\frac{\alpha\pol\,\zre}{\zR\q} \intxyr \xre\s \, I\in (\xre,y,\zre)}}{\displaystyle{\intxyr \left[  \frac{\w\s(\zre)}{\wos} + \frac{\alpha\pol\s\xre\s}{\zR\s}\right] I\in (\xre,y,\zre)}}\,\,.&\nonumber
\end{eqnarray}
By using
\[
\frac{\displaystyle{ \intxyr \xre\s \,I\in (\xre,y,\zre)}}{\displaystyle{ \intxyr\Iref}}=\frac{\w\s(\zre)}{4}\,\,,
\]
we then obtain
\begin{equation}
\label{mean}
\left<\xre\right>\pol \,=\,\frac{2\,\alpha\pol}{(k\,\wo)\q + \alpha\pol\q}\,\,\,\zre\,\,.
\end{equation}
For transverse electric waves  $k\,\wo\gg \alpha\te$ the mean value calculation of angular deviations is 
proportional to $1/(k\,\wo)\q$ and so very small, for the beam parameters used in this work, $k\wo \approx 1.2 \times 10^{^{3}}$. For transverse magnetic waves, in the case of the incidence near the Brewster angle,  
$\theta\bex +\, \delta\,/\,(k\,\wo)$, the factor 
\[\alpha\tm\,\,\to\,\,k\,\wo\,/\,\delta\,\,,\]
and it acts as an amplification

\begin{equation}
\label{meanA}
\left<\xre\right>\tm \,=\,\frac{2\,\delta}{\delta\q + 1}\,\,\,\frac{\zre}{k\,\wo}\,\,.
\end{equation}
From Figure 2, we see that for  $\delta=1$ (i.e. incidence at $\theta\bex+0.05^{^{o}}$) the mean value, due to the presence of a secondary peak,  does not coincide with the maximum of the reflected intensity.  In the next Section, we shall focus our attention on the asymmetry of the reflected beam.

\subsection{The cubic equation}
In order to calculate the angular deviations of the peak of the reflected beam, we equalize to zero  the derivative of  Eq.\,(\ref{Iref}). After simple algebraic manipulations, we find 
\begin{equation}
\label{cubic}
\xre\c +\, a_{\2}^{\pol}\, \xre\s +\, a_{\1}^{\pol}\,\xre = a_{\0}
\end{equation}
where
\begin{eqnarray*}
a_{\2}^{\pol} & = & 2 \,\zre/\,\alpha\pol\,\,,\\
a_{\1}^{\pol} & = &\frac{\w\s(\zre)}{2}\left[2\,\left(\frac{\zR}{\wo\,\alpha\pol}\right)^{^{2}}-1\right]\,\,,
\end{eqnarray*}
and
\[ a_{\0}^{\pol}\, = \,\zre\wzs/\,2\,\alpha\pol\,\,.\]
For transverse electric waves, $k\,\wo\gg \alpha\te$ and this allows to simplify the cubic equation to a linear one 
\[   a_{\1}^{\te}\,\xre^{\mbox{\tiny \,t\,e}} = a_{\0}\,\,,\] 
with $a_{\1}^{\te}=[\w(\zre)\,k\,\wo/2\,\alpha\te]^{^{2}}$, and, consequently, to recover the mean value result
$2\,\alpha\te \zre/(k\,\wo)^{^{2}}$. This shows a symmetry in the reflected (transverse electric) beam. 

Let us now analyse the cubic equation for transverse magnetic waves and incidence near the Brewster angle,
\[ \widetilde{\theta}\,=\,\tbre + \,\delta\,/k\,\wo\,\,.\]
In this case $\alpha\tm \approx k\,\wo/\,\delta$ and the cubic equation becomes  
\begin{equation}
\label{cubB}
\frac{\xre\c}{\woc}\,+\,b_{\2}^{^\mathrm{tm}}\,\frac{\xre\s}{\wos}\,+\,b_{\1}^{^\mathrm{tm}}\,
\,\frac{\xre}{\wo}=b_{\0}^{^\mathrm{tm}}\,\,,
\end{equation}
where
\[
\{\,b_{\2}^{^\mathrm{tm}}\,,\,   b_{\1}^{^\mathrm{tm}}  \,\} \,=\,\{\,\delta\,\zre/\,\zR\,,
(\delta^{^2}-2)\,\w\s(\zre)\,/\,4\,\wos\,\}\,\,,
\]
and
\[b_{\0}^{^\mathrm{tm}} \, = \,\delta\,\zre\,\w\s(\zre)\,/\,4\,\zR\wos\,\,. \]
In Figure 3, we draw the solutions of Eq.\,(\ref{cubB}), continuous lines,  as a function of $\delta$ for different values of $\zre$ (upper plots) and as a function  of  $\zre/\wo$ for different values of $\delta$. The dashed lines
and dots respectively correspond to the mean value and numerical calculation. By decreasing the values of $\delta$ and $\zre$, we amplify the effect of breaking of symmetry and, consequently, the mean value calculation does not represent a good estimate of the angular deviations. The cubic equation solutions show an excellent agreement with the numerical data.

\subsection{The universal solution}

In view of experimental implementations, it should be interesting to find an explicit and, possibly, universal solution of the cubic equation. As seen in Figure 2, for incidence near the Brewster angle an additional peak appears and, in this case, the concept of angular deviations is misleading. Nevertheless, when we increase the axial distance one of the peaks tends to disappear. This observation leads the authors  to consider the limit $\zre\gg\zR$
in the cubic equation (\ref{cubB}). Under such a condition, we can use the approximation 
\[ \w(\zre)\,\approx\,\wo\,\zre\,/\,\zR  \]
and rewrite the coefficients of the cubic equation as follows
\[
\{\,\widetilde{b}_{\2}^{^\mathrm{\,tm}}\,,\,   \widetilde{b}_{\1}^{^\mathrm{\,tm}}  \,\} \,=\,\{\,\delta\,\zre/\,\zR\,,
(\delta^{^2}-2)\,\zre\s\,/\,4\,\zR\s\,\}\,\,,
\]
and
\[\widetilde{b}_{\0}^{^\mathrm{\,tm}} \, = \,\delta\,\zre\c\,/\,4\,\zR\c\,\,. \]
By substituting these coefficients in the cubic equation (\ref{cubB}), we obtain the \textit{universal} equation
\begin{equation}
\label{uniE}
4\,\rho\c\,+\, 4\,\delta\, \rho\s \,+\,
 (\,\delta^{^{2}}-2\,)\,\rho\,=\,\delta\,\,,  
\end{equation}
where
\[\rho\,=\,\frac{\zR\,\xre}{\zre\wo}\,\,. \]
The explicit solution of Eq.\,(\ref{uniE}), which is given by
\begin{equation}
\label{uniS}
\left\{\,-\,\frac{\sqrt{\delta^{^{^{2}}}\,+\,8}\,+\,\delta}{4}\,\,\,,\,\,\,-\,\frac{\delta}{2}\,\,\,,\,\,\,
\frac{\sqrt{\delta^{^{^{2}}}\,+\,8}\,-\,\delta}{4}\,\right\}\,\,,
\end{equation}
contains the information on the two maxima (left and right values) and the minimum (central value). For $\delta=0$, we recover the double peaks phenomenon with the minimum locates at the Brewster angle. For $\delta=1$, the main peak is found at $\rho=1/2$ and this implies 
\[ \xre\, =\, \zre\,/\,k\,\wo\,\,.\]  
This means, in the case of an optical beam with $\wo=100\,\mu\mathrm{m}$ and $\lambda=532\,\mathrm{nm}$, an angular deviation of approximatively $0.05^{^{o}}$ with respect to the optical path predicted by the Snell law with 
an amplification of a factor $k\,\wo$ ($\approx 10^3$) with respect to the angular deviations of transverse 
electric waves.

\section{The (upper) transmitted beam}

The Fresnel coefficient for the (upper) transmitted beam, $T\pol(\kx,\ky)$, contains the double transmission through the left (air/dielectric) and right(dielectric/air) interfaces,
\begin{eqnarray*}
\frac{4\,n\cos\theta\,\cos\psi}{(a\pol\cos\theta+n\cos\psi/a\pol)\q}\,\,,
 \end{eqnarray*}
the reflection at the lower (dielectric/air) interface,
\begin{eqnarray*}
\frac{n\cos\varphi/a\pol-a\pol\cos\phi  }{n\cos\varphi/a\pol+a\pol\cos\phi }\,\,,
 \end{eqnarray*}
and, finally, the optical phase,
\begin{eqnarray*}
 \exp\{\,i\,[\, n\,\cos\varphi\, d\,\sqrt{2}\,+\,(\,n\,\cos\psi\,-\,\cos\theta\,)\,(\,\ell\,-\,d\,)\,]\,\}\,\,,
 \end{eqnarray*}
which takes into account the distance between the the lower and right (dielectric) interfaces and  the left (air/dielectric) one.  This phase is responsible for the optical path predicted by the ray optics\cite{ajp}.

As done for the reflected beam,  we consider the first order Taylor expansion, 
\begin{eqnarray}
T\pol(\kx,\kx)&=& T\pol(0,0)\, \left[\,1\,+\,\beta\pol \,\frac{\kx}{k}\,\right]\,\times \nonumber \\
& & \exp[\,-\,i\,\kx\,x\sn\,]\,\,
\label{TaylorT}
\end{eqnarray}
where
\[
x\sn=(\,\tan\psi\cos\theta\,-\,\sin\theta\,)\,\ell\,+\,(\,\cos\theta\,+\,\sin\theta\,)\,d\,\,,
\]
and
\[\beta\pol=\beta\pol\ua+\beta\pol\ub+\beta\pol\uc\,\,,\]
with
\begin{eqnarray*}
\beta\te\ua&=&\tan\psi\,-\,\tan\theta\,\,,\\
\beta\te\ub&=& 2\,\tan{\phi}\,\cos\theta\,/\,n\,\cos\psi\,\,,\\
\beta\te\uc&=&(\tan\theta\,-\,\tan\psi)\,\cos \theta/\,n\,\cos\psi\,\,,\\
(\sin\s\psi\,-\,\cos\s{\theta})\,\beta\tm\ua&=&\tan\psi\,-\,\tan\theta/\,n\s\,\,,\\
(\sin\s\phi\,-\,\cos\s{\varphi})\,\beta\tm\ub&=&2\,\tan{\phi}\,\cos\theta\,/\,n\,\cos\psi\,\,,\\
(\sin\s\theta\,-\,\cos\s{\psi})\,\beta\tm\uc&=&(\tan\theta\,-\,n\s\,\tan\psi)\,\cos \theta/\,n\,\cos\psi\,\,.
\end{eqnarray*}
The first order expansion of the Fresnel  coefficient allows to analytically solve the integral equation (\ref{Etra}),
\begin{eqnarray}
\label{Etra2}
\Etra
&= &   \left[1 +\,i\,\, \frac{ \beta\pol \xttr +\ztr}{\zR} \right]\times \nonumber \\
& & \hspace*{-1cm}\frac{T\pol ( 0 , 0 ) }{1 + i\, \ztr / \zR}\,\, E\in (\xttr , y , \ztr )\,\,,
\end{eqnarray}
where $\xttr=\xtr-\,x\sn$. 

For the (upper) transmitted beam, before calculating the intensity we have to spend some words on the
$\beta\pol\ub$ coefficients. Two angles play an important role, the internal Brewster angle, 
\[ \sin \phi\bin \,=\, \cos \varphi\bin\,\,\,\,\,\Rightarrow\,\,\,\,\,\, \varphi\bin\,=\,\arctan(1/n)\,\, ,  \]
and the critical one
\[ \cos\phi\cri\,=\,0\,\,\,\,\,\Rightarrow\,\,\,\,\,\, \varphi\cri\,=\,\arcsin(1/n)\,\,.\]
When the incidence angle approaches the internal Brewster or  critical region, the $\beta\tm\ub$  and  $\beta\pol\ub$  coefficients act 
as an amplification factor  of  angular deviations. For
\begin{equation}
\theta\cri\,<\,\frac{1\,-\,\sqrt{n^{^{2}}\,-\,1}}{\sqrt{2}}\,\,,
\end{equation}
the $\beta\pol$ coefficients are real and the (upper) transmitted intensity is then given by
\begin{eqnarray}
\label{Itra}
\Itra&=&  \left[1+ \left( \dfrac{\beta\pol\xttr +\ztr}{\zR} \right)\s\, \right] \times \nonumber \\
& & \hspace*{-2cm} \dfrac{\wos}{\w\s(\ztr)} \,T\pol^{^{2}} ( 0 , 0 )\,\,  I\in (\xttr,y,\ztr)\,\,.
\end{eqnarray}
In both the Brewster angle and the critical one, the coefficient $\beta\ub$ tends to infinity, although for different reasons. In the following sections, the analysis of the angular deviations for these regions is provided.

%%%%%%%%%%%%%%%%%%%%%%%%
%%%%%%%%%%%%%%%%%%%%%%%%
\subsection{The internal Brewster angle}
%%%%%%%%%%%%%%%%%%%%%%%%
%%%%%%%%%%%%%%%%%%%%%%%%
As observed in the previous Section, the internal Brewster angle is found at $\tan\varphi\cri=1/n$. This implies an incidence angle of
\begin{equation}
\theta\bin\,=\,\arcsin\left[\,\frac{n\,(\,1\,-\,n\,)}{\sqrt{\,2\,(\,1\,+\,n^{^{2}})}}\,\right]\,\,.
\end{equation}
At such an incidence, the infinity of the $\beta\tm\ub$ coefficient is compensated by the zero in  $T\tm(0,0)$.
This allows, as done for the reflected beam, to find a closed form for the (upper) transmitted intensity at the internal Brewster angle,
\begin{eqnarray}
\label{IBint}
I\bin\tr(\xtr,y,\ztr) &=&   \left[\,g(n) \,\,\,\dfrac{\xttr}{\zR}\, \dfrac{\wo}{\w\s(\ztr)} \right]\q \times 
\nonumber \\   
 & & I\in (\xttr,y,\ztr)\,\,,
\end{eqnarray}
where
\begin{eqnarray*}
g(n)  &=& \lim_{\,\,\,\,\,\theta\to\theta\bin} \beta\tm T\tm ( 0 , 0 )\\ 
& = & \frac{4\,\sqrt{n\s+1}}{\left(n\,+\,\displaystyle{\frac{1+n}{\sqrt{2+n\s+2\,n\c-n^{\4}}}}\right)^{^{2}}}\,\times\\
& &  \lim_{\,\,\,\,\,\theta\to\theta\bin} \frac{\cos\varphi\,-\,n\,\cos\phi}{\sin\q\phi\,-\,\cos\q\varphi}\\
 & = & \frac{2\,(n^{\4}-1)}{n\,\left(n\,+\,\displaystyle{\frac{1+n}{\sqrt{2+n\s+2\,n\c-n^{\4}}}}\right)^{^{2}}}\,\,.
\end{eqnarray*}
From Eq.\,(\ref{Itra}), we obtain a cubic equation like the one obtained for the reflected beam (\ref{cubic}). The cubic equation that now contains $\beta\pol$ instead of $\alpha\pol$ allows to determine the peak of the upper transmitted beam and consequently the angular deviations. For incidence approaching  the internal Brewster angle,
 $\widetilde{\theta}=\tbri + \delta/k\,\wo$, we find $\beta\tm \approx k\,\wo/\,\delta$ and so we recover Eq.\,(\ref{cubB}) and for axial distance greater than the Rayleigh range Eq.\,(\ref{uniE}). Finally, the reflected and (upper) transmitted beam show the same behaviour for incidence in the Brewster region. The numerical calculation confirms our analytical results.

\subsection{The critical angle}

For incidence approaching  the critical region, both the transverse magnetic and electric $\beta\ub$ coefficient act as angular deviations amplification factors. This is due to the presence in the denominator of $\beta\pol\ub$ of 
$\cos\phi$ which tends to zero when $\theta$ tends to $\theta\cri$. Contrarily to what happens for the Brewster angles this zero in the denominator is not compensated by the zero in the Fresnel coefficient. This implies that, in the critical region, 
\[
\theta\cri\,+\,\delta\,/\,k\,\wo\,\,,
\]
 the first order expansion of the (upper) transmitted beam (\ref{Itra}) and, consequently, the cubic equation obtained for the angular deviation  (\ref{cubic}), with $\beta\pol$ instead of $\alpha\pol$,  is valid for 
$\delta\geq 5$. For these values of $\delta$  the main contribution to the (upper) transmitted beam comes from the real part of the Fresnel coefficient and, in this case, the first order expansion represents a good approximation for the beam.

Figures 4 (for the transverse magnetic waves) and 5 (for the electric ones) show the angular deviations calculated by using the cubic equation  (\ref{cubic}) with $\beta\pol$ in the $a$-coefficients (continuous lines), and the ones obtained by the numerical analysis (dots).  For $\delta\leq -\,5$, the analytical curves and numerical data show an excellent agreement. The plots in Figures 4 and 5 also show an interesting result: For $\delta>-\,5$, the analytical curves show an agreement with the numerical data when we increase the axial distance. This phenomenon clearly deserves future investigation and could shed new light to the field expansion for incidence near the critical angle.

\section{Conclusions}

The first consequence of the symmetry of the wave number distribution  of
a free Gaussian beam is that the mean value and peak position of the optical beam coincide. 
Reflection by and  transmission through a dielectric prism breaks this symmetry. The mathematical
reason is the presence of the Fresnel coefficients in the wave number distributions  of the reflected and transmitted beams. The breaking of symmetry is then responsible for angular deviations from geometric optics.
In this paper, by using the first order Taylor expansion of the Fresnel coefficients, we find an analytic 
expression for the intensity of the optical beam reflected by and (upper) transmitted through a dielectric prism, 
see Eq.\,(\ref{Iref}) and (\ref{Itra}). The closed expressions are then verified by numerical calculations for
incidence near the Brewster angle showing an excellent agreement as can be seen in Figure 2. From these analytical formulas we find a cubic equation which allows to determine the peak of the optical beams and to compare them with the mean value calculation. For transverse magnetic waves, when the incidence angle approaches the Brewster angle, 
$\delta/k\,\wo$, the breaking of symmetry becomes evident and in the limit $\delta=0$   it generates the well-known phenomenon of double peaks. In this region, for an axial distance greater than the   Rayleigh range,  it  is also possible to find a universal solution for the main peak
\[\frac{\zR\,x_{_{\mathrm{peak}}}}{z_{_{\mathrm{beam}}}\wo}\,=\,
\frac{\sqrt{\delta^{^{^{2}}}\,+\,8}\,-\,\delta}{4}\,\,.\]
This simple and universal formula,  explicitly, shows   an  amplification of a factor $k\,\wo$ with respect to the angular deviations  of transverse electric waves which are approximatively of the order of 
$1/ (k\,\wo)^{^{2}}$. It is also important to observe that, notwithstanding the Fresnel coefficients depend on the refractive index, the angular deviations, in this limit, does not  depend on it.
In this paper, it has also been proved that reflected and (upper) transmitted beams show the same behaviour when the incidence approaches the   external and internal Brewster angles. From the previous formula, we immediately see that for $\delta=1$ (in the case analysed in this paper, $\wo=100\,\mu$m and $\lambda=532$ nm, this means an incidence  $0.05^{^{o}}$ around the Brewster angle)  and an axial distance twice the Rayleigh range the shift of the peak with respect to the position predicted by the geometric optics is $\wo$. For the (upper) transmitted beam, when the incidence angle approaches the critical one, angular deviations are amplified both for transverse and for magnetic waves.  The amplifications in the critical region are smaller than the ones seen  in the Brewster region. Finally, angular deviations are also present  in the lower transmitted beam and they represent additional effects to the laser planar trapping recently discussed in literature\cite{LPT}. This topic will be investigated in future studies.

The analytical results found in this paper could be very useful in future experimental studies. In view of possible
experimental implementations, it is important to observe that two kind of experiments can be realized. A first possibility is to us an incident transverse magnetic beam and, then,  to detect the reflected beam by using a camera
positioned parallel to the first prism interface and by increasing its axial distance. Approaching the Brewster angle, we should find deviation from the angular prediction of the refection law.  A second possibility is represented by  mixing transverse electric and magnetic waves. The incidence angle is then chosen near the critical angle.  The  the technique of weak measurements \cite{AD12} allows to further amplify the angular deviations of the
(upper) transmitted beam. For this technique, the closed expressions for the transverse electric and magnetic beams 
presented in this work will play a fundamental role in determining the shift between the peaks of the transmitted beam.

\subsection*{Acknowledgements}

One of the authors (S.D.L.) thanks the CNPq (grant 2018/303911) and
Fapesp (grant 2019/06382-9) for financial support. The authors are also  grateful 
to A. Alessandrelli, L. Maggio, and  L. Solidoro  for their scientific comments and suggestions 
during the preparation of this article and to Profs. G. Co$^\prime$, L. Girlanda, M. Martino,  and M. Mazzeo
for their help in consolidating the research \textit{BRIT} project of international collaboration between 
the State University of Campinas (Brazil) and the Salento University of Lecce (Italy).

\end{document}